\documentstyle[prl,preprint,aps,epsf]{revtex}

\begin{document}
\bibliographystyle{plain}
\title{Newton Law in Brane-World Scenario with $4d$ Induced Gravity:
Singular Quantum Mechanical Approach}
\author{D. K. Park\footnote{e-mail:
dkpark@hep.kyungnam.ac.kr}
} 
\address{
Department of Physics, Kyungnam University, Masan, 631-701, Korea}
\date{\today}
\maketitle

\begin{abstract}
From the viewpoint of the singular quantum mechanics the effect of the 
energy-dependent coupling constant for $\delta$-function potential is 
examined. The energy-dependence of the coupling constant naturally generates
the time-derivative in the boundary condition of the Euclidean propagator.
This is explicitly confirmed by making use of the simple $1d$ model.
The result is applied to the linearized gravity fluctuation equation 
for the brane-world scenario with $4d$ induced gravity.  
Our approach generates $5d$ Newton potential at a certain intermediate
range of distance between two test massive sources. For other range of distance
$4d$ Newton potential is recovered.
\end{abstract}
\maketitle
\newpage
The original brane-world scenarios are designed to solve the gauge-hierarchy
problem, one of the longstanding puzzle in physics, by introducing large
extra dimensions\cite{ark98,ark98-1,ark99} or warped extra 
dimension\cite{rs99-1}.
Especially, Randall-Sundrum(RS) scenario may also provide a mechanism for
the localization of the gravity on the single positive-tension 
brane\cite{rs99-2}. As a result much attention is paid to understand the 
higher-order corrections of Newton potential on the brane by employing the 
various brane-world pictures. Furthermore, these activities may motivate the 
experimental investigation for the short range gravitational 
effect\cite{hoy01}.

For the case of $AdS_5$ bulk with a flat brane as RS scenario the computation
of Newton law carried out in original RS paper\cite{rs99-2} is improved by 
examining the brane-bending effect\cite{garr00,gidd00} or by computing
the one-loop corrections to the gravitational propagator\cite{duff00}. 
These improvements introduce a mutiplication factor in the subleading term
of Newton potential
contributed from Kaluza-Klein excitations. In addition, the linearized 
gravity 
fluctuation equation is also treated from the viewpoint of the singular
quantum mechanics(SQM)\cite{park01,park02}. In SQM it is well-known that 
when the potential is too singular, Hamiltonian loses its self-adjoint 
property, and thus the conservation of the probability met a serious
problem.
In order to overcome this difficulty we should maintain the self-adjoint
property of Hamiltonian by extending its domain of definition appropriately,
which is refered to a self-adjoint extension\cite{cap85,reed75}. It is known 
that this mathematical approach is effectively identical with the 
physically-oriented
coupling constant renormalization scheme\cite{jack91,park95}. The method has 
been applied to the gravitation fluctuation equations of RS single
brane and two brane scenarios for the compromise of the gravitational 
localization with a small cosmological 
constant\cite{park01,park02}. It also generates the logarithmic correction
in the short range of Newton potential\cite{park03}.

Newton law with a different setup is also examined. The gravitational 
potential for the flat brane in $dS_5$ bulk and for the curved $dS$ brane in 
$dS_5$ or $AdS_5$ bulk are examined\cite{noji02}. For the case of the flat
brane in $dS_5$ bulk the sign of the subleading correction is changed to be
negative. The physical implication of this sign change is discussed in
Ref.\cite{noji02} in the context of $dS$/CFT correspondence\cite{strom01}.

Another type of the scenario which attracts an attention recently is a 
brane-world with a $4d$ induced Einstein term where the brane has its
own gravity term {\it ab initio}. For the case of the flat bulk in this 
picture the gravitational potential becomes $4d$ type $1/r$ at the short
range, {\it i.e.} $r << \lambda / 2$ and 5d type $1 / r^2$ at the long
range, {\it i.e.} $r >> \lambda / 2$, where $\lambda$ is a ratio of $4d$
Planck scale with that of $5d$: $\lambda \equiv M_4^2 / M_5^3$\cite{dvali00}.
This fact can be used for explaining the acceleration of the 
universe\cite{ruba03}.
Newton law with $4d$ induced gravity in the $AdS_5$ background is also 
examined\cite{kiri02,ito02}. In this case there is an intermediate range
of distance where $5d$ gravitational potential plays a dominant role.
At other ranges $4d$ gravitational potential is recovered.

The most remarkable feature of the linearized fluctuation equation for the
case of $AdS_5$ bulk with the induced Einstein term is the fact that 
the coupling constant of the $\delta$-function potential is dependent on 
energy(or mass) as following:
\begin{equation}
\label{fluct1}
\hat{H} \hat{\psi}(z) = E \hat{\psi}(z)
\end{equation}
where
\begin{eqnarray}
\label{hamil1}
\hat{H}&=& \hat{H}_{RS} - \lambda E \delta(z)   \\  \nonumber
\hat{H}_{RS}&=& -\frac{1}{2} \frac{\partial^2}{\partial z^2} + 
\frac{15 k^2}{8 (k |z| + 1)^2} - \frac{3}{2} k \delta(z)
\end{eqnarray}
and $\lambda \equiv M_4^2 / M_5^3$. If we regard Eq.(\ref{fluct1}) as a 
Schr\"{o}dinger equation, we should rely on the SQM with an energy-dependent 
coupling constant of $\delta$-function potential. 
Upon our knowledge the effect of the energy-dependent coupling constant was
not fully examined in the context of SQM. 
The purpose of this letter
is to analyze Eq.(\ref{fluct1}) from the viewpoint of SQM.

To understand the effect of the energy-dependent coupling constant in 
quantum mechanics let us consider the Schr\"{o}dinger equation
$H \phi = E \phi$, where
\begin{eqnarray}
\label{hamil2}
& &H = H_V(\vec{p}, \vec{r}) + \hat{v}(E) \delta(\vec{r})  \\  \nonumber
& &H_V(\vec{p}, \vec{r}) = \frac{\vec{p}^2}{2} + V(\vec{r}).
\end{eqnarray}
Let $G[\vec{r}_1, \vec{r}_2; t]$ and $G_V[\vec{r}_1, \vec{r}_2; t]$ be the
Euclidean propagators for $H$ and $H_V$ respectively. Then, it is well-known
that $G[\vec{r}_1, \vec{r}_2; t]$ satisfies the following integral equation
\begin{equation}
\label{integral}
G[\vec{r}_1, \vec{r}_2; t] = G_V[\vec{r}_1, \vec{r}_2; t] - 
\hat{v}(E) \int_0^t ds G_V[\vec{r}_1, \vec{0}; t - s] G[\vec{0}, \vec{r}_2; s].
\end{equation}
If we take a Laplace transform
\begin{equation}
\label{laplace}
{\cal L}[f] \equiv \hat{f}(E) 
\equiv \int_0^t dt f(t) e^{-E t}
\end{equation}
in Eq.(\ref{integral}), one can easily derive
\begin{equation}
\label{relation1}
\hat{G}[\vec{r}_1, \vec{r}_2; E] = \hat{G}_V[\vec{r}_1, \vec{r}_2; E]
- \hat{v}(E) \hat{G}_V[\vec{r}_1, \vec{0}; E] \hat{G}[\vec{0}, \vec{r}_2; E].
\end{equation}
Solving Eq.(\ref{relation1}) one can express the fixed-energy amplitude
$\hat{G}[\vec{r}_1, \vec{r}_2; E]$ from $\hat{G}_V[\vec{r}_1, \vec{r}_2; E]$;
\begin{equation}
\label{fixed-e1}
\hat{G}[\vec{r}_1, \vec{r}_2; E] = \hat{G}_V[\vec{r}_1, \vec{r}_2; E]
- \frac{\hat{G}_V[\vec{r}_1, \vec{0}; E] \hat{G}_V[\vec{0}, \vec{r}_2; E]}
       {\frac{1}{\hat{v}(E)} + \hat{G}_V[\vec{0}, \vec{0}; E]}.
\end{equation}

As an example let us consider a simple $1d$ free particle case for $H_V$.
Then, the fixed-energy amplitude for $H_V$ is simply
\begin{equation}
\label{free1}
\hat{G}_V[x, y; E] \equiv \hat{G}_F[x, y; E] = 
\frac{e^{-\sqrt{2E} |x - y|}}{\sqrt{2E}}
\end{equation}
where the subscript `F' stands for the free particle.
Thus, Eq.(\ref{fixed-e1}) can be re-written as
\begin{equation}
\label{fixed-e2}
\hat{G}[x, y; E] = \frac{e^{-\sqrt{2E} |x - y|}}{\sqrt{2E}} - 
\frac{\hat{v}(E)}{\sqrt{2E} ( \sqrt{2E} + \hat{v}(E) )}
e^{-\sqrt{2E} (|x| + |y|)}.
\end{equation}

If $\hat{v}(E)$ is independent of $E$, one can easily take an inverse Laplace
transform to Eq.(\ref{fixed-e2}) using the formulae
\begin{eqnarray}
\label{inverse-lap}
{\cal L}^{-1} \left( \frac{1}{\sqrt{E}} e^{-\alpha^{\frac{1}{2}} 
                                            E^{\frac{1}{2}} } \right)
&=& \pi^{-\frac{1}{2}} t^{-\frac{1}{2}} e^{-\frac{\alpha}{4 t}}
                                                     \\   \nonumber
{\cal L}^{-1} \left( (\sqrt{E} + \beta)^{-1} e^{-\alpha E^{\frac{1}{2}}} \right)
&=& \pi^{-\frac{1}{2}} t^{-\frac{1}{2}} e^{-\frac{\alpha^2}{4 t}} - 
\beta e^{\alpha \beta + \beta^2 t}
Erfc \left( \frac{1}{2} \alpha t^{-\frac{1}{2}} + \beta t^{\frac{1}{2}}
                                                                \right)
\end{eqnarray}
where $Erfc(z)$ is an usual error function:
\begin{equation}
\label{error}
Erfc(z) \equiv \frac{2}{\sqrt{\pi}} 
\int_z^{\infty} e^{-t^2} dt.
\end{equation}
After some manipulations one can show the Euclidean propagator in this case is 
simply
\begin{equation}
\label{eucli1}
G[x, y; t] = G_0[x, y; t] - \hat{v} \int_0^{\infty} dz
e^{-\hat{v} z} G_0[|x|, -|y|-|z|; t]
\end{equation}
where $G_0[x, y; t]$ is a propagator for a $1d$ free-particle system;
\begin{equation}
\label{free2}
G_0[x, y; t] = \frac{1}{\sqrt{2 \pi t}} e^{-\frac{(x - y)^2}{2 t}}.
\end{equation}
One can show that $G[x, y; t]$ and $\hat{G}[x, y; E]$ satisfy the usual
boundary condition for the $\delta$-function potential;
\begin{eqnarray}
\label{boundary1}
\frac{\partial G}{\partial x} [0^+, y; t] - 
\frac{\partial G}{\partial x} [0^-, y; t]&=& 2 \hat{v} G[0, y; t]
                                                        \\   \nonumber
\frac{\partial \hat{G}}{\partial x} [0^+, y; E] - 
\frac{\partial \hat{G}}{\partial x} [0^-, y; E]&=&
2 \hat{v} \hat{G}[0, y; E].
\end{eqnarray}

Next, let us consider $\hat{v}(E) = \alpha E$ case. Then, Eq.(\ref{fixed-e2})
shows that the fixed-energy amplitude $\hat{G}[x, y; E]$ becomes
\begin{equation}
\label{fixed-e3}
\hat{G}[x, y; E] = \frac{e^{-\sqrt{2E} |x - y|}}{\sqrt{2E}}
- \frac{1}{\sqrt{2}}
\frac{1}{\sqrt{E} + \frac{\sqrt{2}}{\alpha}}
e^{-\sqrt{2E} (|x| + |y|)}.
\end{equation}
From Eq.(\ref{fixed-e3}) one can show $\hat{G}[x, y; E]$ satisfies the 
following boundary condition at $x = 0$;
\begin{equation}
\label{boundary2}
\frac{\partial \hat{G}}{\partial x} [0^+, y; E] - 
\frac{\partial \hat{G}}{\partial x} [0^-, y; E] = 2 \hat{v}(E)
\hat{G}[0, y; E].
\end{equation}
Eq.(\ref{boundary2}) enables us to derive a boundary condition for the 
corresponding 
Euclidean propagator $G[x, y; t]$ to be satisfied at $x = 0$. Taking an 
inverse Laplace transform in Eq.(\ref{boundary2}) leads to 
\begin{equation}
\label{relation2}
\frac{\partial G}{\partial x} [0^+, y; t] - 
\frac{\partial G}{\partial x} [0^-, y; t] = 2 \int_0^t ds v(t - s) 
G[0, y; s]
\end{equation}
where $v(t) = {\cal L}^{-1} [\hat{v}(E)]$.
Since $v(t)$ becomes 
\begin{equation}
\label{time-coupl}
v(t) = \alpha \lim_{\epsilon \rightarrow 0^+} \delta^{\prime}(t - \epsilon)
\end{equation}
for $\hat{v}(E) = \alpha E$, Eq.(\ref{relation2}) reduces to 
\begin{equation}
\label{boundary3}
\frac{\partial G}{\partial x} [0^+, y; t] - 
\frac{\partial G}{\partial x} [0^-, y; t] = 2 \alpha 
\frac{\partial}{\partial t} G[0, y; t].
\end{equation}
Thus the energy-dependence of the coupling constant yields an time-derivative
at the boundary condition of the propagator. This may be understood
from the usual energy-time uncertainty principle. The explicit form of 
the Euclidean
propagator in this case can be derived by taking an inverse Laplace transform
to Eq.(\ref{fixed-e3});
\begin{equation}
\label{eucli2}
G[x, y; t] = G_0[x, y; t] - G_0[|x|, -|y|; t] + 
\frac{2}{\alpha} \int_0^{\infty} dz e^{-\frac{2}{\alpha} z}
G_0[|x|, -|y|-|z|; t]
\end{equation}
where $G_0[x, y; t]$ is given in Eq.(\ref{free2}). It is straightforward to 
show that $G[x, y; t]$ really satisfies the boundary condition 
(\ref{boundary3}). 

Now, let us go back to the gravitational fluctuation equation (\ref{fluct1}).
Firstly, let us comment briefly how Eq.(\ref{fluct1}) is derived. The $5d$
Einstein equation we consider is 
\begin{equation}
\label{einstein1}
\left(\tilde{R}_{MN} - \frac{1}{2} G_{MN} \tilde{R} \right)
+ \frac{\Lambda}{M_5^3} G_{MN}
+ \lambda \left[ \left(R_{\mu \nu} - \frac{1}{2} g_{\mu \nu} R\right)
                + \frac{v_b}{M_4^2} g_{\mu \nu} \right] \delta_M^{\mu}
                 \delta_N^{\nu} \delta(y) = 0
\end{equation}
where $\tilde{R}$ and $R$ are $5d$ and $4d$ curvature scalar respectively.
Of course $\Lambda$ and $v_b$ are $5d$ cosmological constant and brane 
tension. In fact, Eq.(\ref{einstein1}) can be derived by varying the 
Einstein-Hilbert action
\begin{equation}
\label{action1}
S = \int d^x dy \sqrt{-G}
\left[ \frac{M_5^3}{2} \tilde{R} - \Lambda + 
       \left( \frac{M_4^2}{2} R - v_b \right) \delta(y) \right].
\end{equation}
The solution of Eq.(\ref{einstein1}) we have interest is same with that of the
usual RS scenario
\begin{equation}
\label{solution1}
ds^2 = e^{-2 k |y|} \eta_{\mu \nu} dx^{\mu} dx^{\nu} + dy^2
\end{equation}
with the fine-tuning conditions $\Lambda = -6 M_5^3 k^2$ and $v_b = 6 k M_5^3$.
The coincidence of the classical solution with that of the usual RS scenario
is in fact evident because the 
$3$-brane in Eq.(\ref{solution1}) is flat and hence the $4d$ induced gravity 
does not play any role. However, this induced gravity yields an important
effect when we consider the linearized gravitational fluctuation 
$h_{\mu \nu}(x, y)$ defined as 
\begin{equation}
\label{linearized}
ds^2 = (e^{-2 k |y|} \eta_{\mu \nu} + h_{\mu \nu}) dx^{\mu} dx^{\nu} + dy^2.
\end{equation}
Inserting Eq.(\ref{linearized}) into the Einstein equation (\ref{einstein1})
with redefinition $h_{\mu \nu} = e^{-k |y| / 2} \hat{\psi}(y) e^{ipx}$,
$z = \epsilon(y) (e^{k|y|} - 1) / k$, and $E = -p^2 / 2$, one can derive
Eq.(\ref{fluct1}) straightforwardly. Of course we should use RS gauge
$h^{\nu}_{\mu, \nu} = h^{\mu}_{\mu} = 0$ in the course of calculation.

The fixed-energy amplitude $\hat{G}_{RS}[a, b; E]$ for $\hat{H}_{RS}$ is 
examined in detail in Ref.\cite{park01,park02}. In fact, 
$\hat{G}_{RS}[a, b; E]$ is crucially dependent on the boundary condition
at $x = R \equiv 1/k$, where $x = z + R$, which is parametrized by $\xi$ in
Ref.\cite{park01,park02}. Here, we choose only $\xi = 1/2$ case which means the
Dirichlet and the Neumann boundary conditions are included with equal weight.
In this case the fixed-energy amplitude $\hat{G}_{RS}[a, b; E]$ on the brane
simply reduces to 
$\hat{G}_{RS}[a, b; E] = \Delta_0^{RS} + \Delta_{KK}^{RS}$ where
\begin{equation}
\label{zerokk}
\Delta_0^{RS} = \frac{1}{R E}
\hspace{1.0cm}
\Delta_{KK}^{RS} = \frac{1}{\sqrt{2 E}}
\frac{K_0(\sqrt{2 E} R)}{K_1(\sqrt{2 E} R)}
\end{equation}
and $K_{\nu}(z)$ is an usual modified Bessel function. Of course, 
$\Delta_0^{RS}$ and $\Delta_{KK}^{RS}$ represent the zero mode and  
Kaluaza-Klein excitations of the RS scenario respectively. 

The general formula Eq.(\ref{fixed-e1}) enables us to derive a 
fixed-energy amplitude
$\hat{G}[a, b; E]$ for Hamiltonian $\hat{H}$ defined in Eq.(\ref{hamil1})
as following:
\begin{equation}
\label{fixed-e4}
\hat{G}[a, b; E] = \hat{G}_{RS}[a, b; E] + 
\frac{\hat{G}_{RS}[a, R; E] \hat{G}_{RS}[R, b; E]}
     {\frac{1}{\lambda E} - \hat{G}_{RS}[R, R; E]}.
\end{equation}
Thus, the fixed-energy amplitude at the location of the brane for the system
involving the $4d$ induced gravity is 
\begin{equation}
\label{fixed-e5}
\hat{G}[R, R; E] = (\Delta_0^{RS} + \Delta_{KK}^{RS}) + 
\frac{(\Delta_0^{RS} + \Delta_{KK}^{RS})^2}
     {\frac{1}{\lambda E} - (\Delta_0^{RS} + \Delta_{KK}^{RS})}
=
\frac{2}{m} 
\frac{K_2(m R)}{2 K_1(m R) - \lambda m K_2(m R)}
\end{equation}
where $m \equiv \sqrt{2 E}$.
The fixed-energy we derived in Eq.(\ref{fixed-e5}) is proportional to the
momentum-dependent Green function $\tilde{G}_R(p, y=0)$ which is expressed
at Eq.(2.10) of Ref.\cite{kiri02}. If we adjust our conventions with those
of Ref.\cite{kiri02}, the relation between them is simply
$\hat{G}[R, R; E] = M_5^3 \tilde{G}_R(p = m, y=0)$. This simple relation makes
us to analyze Newton law of the gravitation localized on the brane.

Newton potential localized on the brane is generally obtained from the 
space-time dependent Green function as following\cite{garr00,dvali00}
\begin{equation}
\label{rev1}
V(\vec{x}) = \int dt G_R(t, \vec{x}, y=0; 0, \vec{0}, 0)
\end{equation}
where the subscript `R' stands for the retarded Green function.
Using a Fourier transform of $G_R$
\begin{equation}
\label{rev2}
G_R(t, \vec{x}, y; 0, \vec{0}, 0) \equiv \int \frac{d^4 p}{(2 \pi)^4}
e^{i p x} \tilde{G}_R(p, y)
\end{equation}
one can show the potential $V(\vec{x})$ in Eq.(\ref{rev1}) reduces to 
\begin{equation}
\label{rev3}
V(r) = \frac{1}{2 \pi^2 r} \int_0^{\infty} dp p \sin pr
\tilde{G}_R(p, y=0)
= \frac{1}{2\pi^2 M_5^3 r} \int_{m_0}^{\infty} dm \sin mr
\hat{G}[R, R; E].
\end{equation} 
Since the continuum states start from the asymptotic value of the 
quantum-mechanical potential, the singular $\delta$-function potential can
not generate any mass gap. Thus we can conclude $m_0 = 0$ in Eq.(\ref{rev3}).
The factor $\sin mr$ in Eq.(\ref{rev3}) is crucial to extract 
an information on the long-range
and short-range behaviors of the gravitational potential $V(r)$. To show
this more explicitly we re-express $V(r)$ as following
\begin{equation}
\label{rev4}
V(r) = \frac{1}{\pi^2 M_5^3 R r} \int_0^{\infty} du 
\sin \left( \frac{r}{R} u \right)
\frac{K_2(u)}{2 K_1(u) - \frac{\lambda}{R} u K_2(u)}.
\end{equation}

If $r >> R$, the high oscillation of $\sin(r u/R)$ results in a negligible
contribution to the integral from large $u$. Thus, one can approximate
$K_1(u)$ and $K_2(u)$ as
\begin{equation}
\label{rev5}
K_1(u) \sim \frac{1}{u} + \frac{u}{2} \ln u,
\hspace{1.0cm}
K_2(u) \sim \frac{1}{u^2} - \frac{1}{2}.
\end{equation}
Then it is straightforward to show that the potential becomes
\begin{equation}
\label{rev6}
V(r) \sim \frac{1}{2\pi M_5^3 R \left( 2 - \frac{\lambda}{R} \right) r}
\left[ 1 - \frac{2}{\pi \left(2 - \frac{\lambda}{R} \right)}
\lim_{\epsilon \rightarrow 0^+}
\int_0^{\infty} du u e^{-\epsilon u} \sin \left( \frac{r}{R} u \right)
\ln u \right]
\end{equation}
where the infinitesimal parameter $\epsilon$ is introduced for the 
regularization. Performing the integration in Eq.(\ref{rev6}) makes
$V(r)$ to be
\begin{equation}
\label{rev7}
V(r) \sim \frac{1}{2\pi M_5^3 R \left( 2 - \frac{\lambda}{R} \right) r}
\left( 1 + \frac{1}{2 - \frac{\lambda}{R}} \frac{R^2}{r^2} \right).
\end{equation}
Thus the potential exhibits the $4d$ behavior at the long-range.

If $r << R$, the large $u$ region mainly contributes to the integral of 
Eq.(\ref{rev4}). Thus, one can use the asymptotic expansion of 
$K_1(u)$ and $K_2(u)$;
\begin{equation}
\label{rev8}
K_1(u) \sim \sqrt{\frac{\pi}{2 u}} e^{-u} 
            \left( 1 + \frac{3}{8 u} \right),
\hspace{1.0cm}
K_2(u) \sim \sqrt{\frac{\pi}{2 u}} e^{-u}
            \left( 1 + \frac{15}{8 u} \right).
\end{equation}
Then it is straightforward to compute $V(r)$ whose final approximate expression
is 
\begin{equation}
\label{rev9}
V(r) \sim -\frac{1}{\pi^2 M_5^3 \lambda r}
\left[ \frac{\pi}{2} - 
\frac{\frac{2R}{\lambda}}{\frac{2R}{\lambda} - \frac{15}{8}}
\left\{\frac{\pi}{2} - f\left(\frac{r}{R} \left(\frac{2R}{\lambda} - 
\frac{15}{8} \right) \right) \right\} \right]
\end{equation}
where $f(z) \equiv ci(z) \sin z - si(z) \cos z$ and, $ci(z)$ and $si(z)$ are 
usual sine and cosine integral functions.
If $\lambda << R$, $V(r)$ simply reduce to 
\begin{equation}
\label{rev10}
V(r) \sim -\frac{1}{\pi^2 M_5^3 \lambda r}
f\left(\frac{2 r}{\lambda} \right).
\end{equation}
Thus if $\lambda << r$, one can use the asymptotic expansion of 
$f(z)$\cite{milton} 
\begin{equation}
\label{rev11}
\lim_{z \rightarrow \infty} f(z) \sim \frac{1}{z}
\left( 1 - \frac{2}{z^2} \right)
\end{equation}
which results in
\begin{equation}
\label{rev12}
V(r) \sim -\frac{1}{2\pi^2 M_5^3 \lambda r^2}
\left( 1 - \frac{\lambda^2}{2 r^2} \right).
\end{equation}
Therefore, in the region $\lambda << r << R$, $V(r)$ displays the $5d$ 
gravitational
behavior. If $r << \lambda << R$, Eq.(\ref{rev10}) shows $V(r)$ reduces to
\begin{equation}
\label{rev13}
V(r) \sim -\frac{1}{2 \pi M_5^3 \lambda r}
\left( 1 + \frac{4 r}{\pi \lambda} \ln \frac{2 r}{\lambda} \right).
\end{equation}
Thus at this region $V(r)$ recovers the $4d$ behavior.

The exact computation of Newton potential in the full range of $r$ with 
arbitrary $\lambda$ seems to be interesting. It may need an highly nontrivial
numerical method because our computation requires a regularization. 
We hope to return to this issue in the near future.
Another point we should stress
is that our result comes from 
$\hat{G}_{RS}[R, R; E] = \Delta_0^{RS} + \Delta_{KK}^{RS}$. In 
Ref.\cite{park01,park02}, however, $\hat{G}_{RS}[a, b; E]$ is crucially
dependent on the boundary conditions, which is parametrized by $\xi$. It is 
interesting to study an effect of the induced gravity when the general 
boundary conditions are employed. From the viewpoint of SQM
Eq.(\ref{fixed-e1}) is a formal solution for the fixed-energy amplitude
because $\hat{G}_V[\vec{r}, \vec{0}; E]$ is in general ill-defined at the higher
dimensional theory. Thus one should adopt an appropriate regularization scheme
for obtaining the finite result\cite{jack91,park95}. It seems to be interesting
to study further the effect of the energy-dependent coupling constant from
the aspect of pure SQM.

\vspace{1.0cm}

{\bf Acknowledgement:} 
This work was supported by the Kyungnam University
Research Fund, 2002.







\end{document}